\documentclass[a4paper,11pt]{article}

\usepackage{jcappub} 

\usepackage[T1]{fontenc} 

\usepackage{siunitx}
\usepackage{myaasmacros}
\newcommand{\be}{\begin{equation}}
\newcommand{\ee}{\end{equation}}
\newcommand{\mr}{\mathrm}

\title{Constraints on the AGN flares as sources of ultra-high energy cosmic rays
from the Fermi-LAT observations}

\author[a,1]{B.A. Nizamov,\note{Corresponding author.}}
\author[a,b,c]{M.S. Pshirkov}

\affiliation[a]{Sternberg Astronomical Institute, Lomonosov Moscow State
University,\\Universitetsky prospekt 13, 119992, Moscow, Russia}
\affiliation[b]{Institute for Nuclear Research of the Russian Academy of Sciences,\\60th October Anniversary st. 7a, 117312, Moscow, Russia}
\affiliation[c]{P.N. Lebedev Physical Institute, Pushchino Radio Astronomy
Observatory,\\142290 Pushchino, Russia}

\emailAdd{nizamov@physics.msu.ru}
\emailAdd{pshirkov@sai.msu.ru}

\abstract{It is well known that if the most energetic cosmic rays ($E>10^{20}$~eV) were
protons then their acceleration sites should possess some extreme properties,
including gigantic luminosity. As no stationary sources with such properties are
known in the local ($D<200$~Mpc) neighborhood of the Milky Way, it is highly
likely that the UHECR acceleration takes place in some transient events. In this
paper we investigate scenario where the UHECRs are produced in strong AGN
flares. Using more than 7 years of the Fermi-LAT observations we select
candidate flares and, using correlation between jet kinetic luminosity and its
bolometric luminosity, estimate local kinetic emissivity of giant AGN flares:
$\mathcal{L}\sim3.7\times10^{44}\mathrm{erg~Mpc^{-3}~{yr^{-1}}}$. This value is
about an order of magnitude larger than the emissivity in the CRs with
$E>10^{20}$~eV, thus making this scenario  feasible, if the UHECR escape
spectrum in these sources is rather hard and/or narrow. This shape of spectrum is predicted in a number of present models of strong relativistic collisionless shocks. Also the scenario of acceleration in AGN flares can accommodate  constraints coming from the observed arrival distribution of UHECRs. Finally, we demonstrate that in case of heavier UHECR composition all the constraints are greatly relaxed.}

\begin{document}
\maketitle
\flushbottom

\section{Introduction}\label{sec:intro}
Despite huge efforts to identify sources of ultra-high energy cosmic rays (UHECR,
$E>10^{18}~$eV), the problem of their origin is  far from being resolved --
we do not know where these
particles are accelerated \cite{Letessier-Selvon2011}. Pure common sense
suggests that the acceleration to such extreme energies takes place in regions
with some extreme conditions,  and it can be demonstrated more rigorously  using
so-called ``Hillas plot'' \cite{Hillas1984}: acceleration of  energetic
particles requires  non-trivial combination of source parameters, primarily
magnetic field strength $B$ and acceleration region size $L$ and that severely
decreases the number of potential sources (e.g.,  \cite{Ptitsyna2010}).
Active galactic nuclei (AGNs) are considered to be one of the most plausible
candidates   -- it is very tempting to tap into a huge reservoir of energy
coming from accretion of matter onto supermassive black hole. However, this
model faces  certain difficulties -- a  configuration needed for an effective
acceleration is a very efficient photon source   as well. It can be estimated
that sources that can accelerate protons up to the highest energies exceeding
$10^{20}$~eV will have a  luminosity around 
$10^{47}~\mathrm{erg~s^{-1}}$ \cite{Lovelace1976,Blandford2000,Farrar2009,Waxman2009}. On the other hand,
production of observed UHECR is a local phenomenon, because due to an
interaction with background photons the horizon of propagation of $10^{20}$~eV
protons is smaller than 150 Mpc \cite{Greisen1966,Zatsepin1966}. The existence
of Greisen-Zatsepin-Kuzmin (GZK) cut-off in the UHECR energy spectrum was
discovered by the HiRes experiment \cite{HiRes_GZK} and lately confirmed with a
very high statistical significance by the Pierre Auger and the Telescope Array
observatories \cite{PA_GZK,TA_GZK}. The problem is rather self-evident --- 
there are no steady sources with required luminosity in the local volume.
It was suggested that this difficulty can be solved by dropping the 'steadiness'
condition --  UHECRs can be accelerated in flares of AGNs. Flares with isotropic
luminosity $>10^{50}~\mathrm{erg~s^{-1}}$ were observed \cite{AGN_flare}, so it
is theoretically possible to accelerate protons up to energies exceeding
$10^{21}$~eV\footnote{One should bear in mind that the cut-off in the
spectrum can be explained by the extinction of sources with that extreme 
parameters rather than by interaction with the CMB photons.}.

This model can be tested observationally. In order to do that we have used
$\gamma$-ray observations of the Fermi-LAT in the high energy (HE, $E>100$~MeV)
range. Fermi-LAT observes the whole sky every three  hours since August 2008,
providing almost uniform coverage with a high temporal resolution. We have
focused on the 'local sources' ($z<0.3$) because we wanted to avoid
complications that arise due to  a redshift evolution but still retain a decent
statistics, with a probed volume that is 1000 times larger than a GZK
volume.

We have selected all flares with a luminosity above the threshold corresponding to
acceleration of UHECR protons to energies $E>10^{20}~$eV and calculated total
fluence of these flares. That allowed us to obtain local emissivity in the HE
$\gamma$-rays,  estimate the kinetic emissivity, and finally
obtain the ratio of UHECR to kinetic emissivity.

The paper is organized as follows: in  Section~\ref{sec:met} we describe the
data and method of data analysis, Section~\ref{sec:dis} contains our results and
discussion, and we draw conclusions in Section~\ref{sec:conclusion}.

\section{Data and method}\label{sec:met}
The goal of the present work is to investigate whether AGNs can be possible
sources of UHECRs with $E>10^{20}$~eV even in the case of the lightest
(protonic) composition. Only sources with some extreme properties can accelerate
protons to these energies \cite{Hillas1984, Ptitsyna2010} and the closest
stationary source of the kind resides far out the GZK volume $V_{\mathrm{GZK}}$\footnote{Selection of $V_\mathrm{GZK}$ is meaningful even in the case of `dying sources' scenario, because GZK horizon is set by the interaction of UHECRs with the photon background and does not depend on the properties of the sources},
which we define as a sphere with a radius $R_{\mathrm{GZK}}=150~$Mpc -- a  mean
attenuation free path of a particle with the \textit{initial} energy of
$10^{20}$ eV. However, strong flares of AGNs can possibly accelerate protons to
the very highest energies \cite{Farrar2009,Waxman2009}.
First, the threshold in luminosity which corresponds to the cosmic ray energy
$E>10^{20}$~eV shall be defined. Let us consider a spherical blob within a jet.
We assume that both particle acceleration and emission of the radiation take
place in this blob. Let its radius and the magnetic field strength be $R', B'$
where the primes indicate the comoving reference frame. The stringent
constraints on the size of the acceleration region and strength of the magnetic
field \cite{Hillas1984} can be written out as follows:
\begin{equation}
B'R' \geq E'/Ze, \label{eq:hillas}
\end{equation}
where $E'$ is the UHECR energy in the comoving frame, $Z$ is the charge of the
particle (for the moment, we consider protons, but we keep $Z$ for generality).
Next, let us connect the
radiative luminosity $L_\mr{r}$ (by which we mean the synchrotron luminosity) to
the energy of UHECR. Radiative luminosities in the observer's and
comoving frames are related as $L_\mr{r} = L'_\mr{r} \Gamma^4$. If the radiation
energy density in the blob is $U'_\mr{r}$ then the flux from the unit surface
area is $(c/4)U'_\mr{r}$ and the luminosity of the spherical blob is
$L'_\mr{r} = \pi R'^2 c U'_\mr{r}$, or, assuming energy equipartition between
radiation and magnetic field,
$L'_\mr{r} = \pi R'^2 c U'_\mr{b} = \pi R'^2 c B'^2/8\pi \geq (c/8)(E'/Ze)^2$.
Since $E' = E/\Gamma$, we finally obtain
\begin{equation}
L_\mr{r} \geq \frac{c}{8e^2} E^2 \frac{\Gamma^2}{Z^2} \approx 4 \times 10^{44} E_{20}
\frac{\Gamma^2}{Z^2} \text{erg/s}	\label{eq:maglum_thr}
\end{equation}
which is adopted in this paper as a threshold value; $E_{20}$ is the UHECR
energy in units $10^{20}$ eV.

Equating magnetic and radiative energy densities is justified by the fact that
the ratio of the luminosities at the two spectral peaks in SED is equal to the
ratio of the synchrotron and magnetic energy densities (Eq.~5 of
\cite{Tavecchio1998}) and typically, heights of two peaks in a BL Lac object's
SED are comparable\footnote{Modeling presented in, e.g., \cite{Ghisellini2005,
Inoue2016} also shows that the two energy densities are close to each other, on
average.} (it will be seen later that our objects of interest are mostly
BL Lacs). The same circumstance allows us to use $L_\gamma$, the high-energy
luminosity in the range 0.1--100 GeV, as a reliable proxy of $L_\mr{r}$.

Once again, we try to link the maximal UHECR energy to an observable quantity
through several steps: 1) get  constraints on   the magnetic field strength and
the size of acceleration region via Hillas criterion, 2) link magnetic and
radiative synchrotron energy densities on the basis of energy equipartition and
observed SEDs, 3) finally, estimate the  synchrotron luminosity from the
gamma-ray luminosity using the properties of  BL Lac SEDs. More discussion on
these relations will be given in Section~\ref{sec:dis:models}.

Note that in principle,
energy gain can take place in so-called dark accelerators without considerable
emission from leptons, but such models should realize very specific conditions,
like linear accelerators in vicinity of rotating dormant black holes \cite{Boldt1999}. The magnetic
field must be always parallel to the velocity of the leptons and this condition
is not satisfied in usual stochastic Fermi acceleration taking place in the jets
of the AGNs -- one can not avoid abundant radiation from accelerated particles,
especially leptons in this less ordered environments. Another possibility of 'dark acceleration' arises in electron-starved environments: if $\epsilon_e$ was low then the radiation would be naturally suppressed. It is not easy to realize scenario of the kind in usual  astrophysical conditions, particularly in ones existing in the AGN jets.

Flares satisfying the condition (\ref{eq:maglum_thr}) have never been observed
in $V_\mr{GZK}$. Nevertheless it is possible to calculate their local energetics
using much larger test volume $V_0$: we have selected a sphere with a radius
corresponding to $z=0.3, ~V_0 \sim 10^{3}V_{\mathrm{GZK}}$. This volume is large
enough to avoid significant  statistical fluctuations and still it is
appropriate to neglect effects of cosmological evolution at $z<0.3$. We selected
all flaring sources satisfying Eq. (\ref{eq:maglum_thr}), that potentially could
be the sites of UHECR acceleration.

For almost 10 years Fermi-LAT has been continuously observing the celestial
sphere at energies $>100$~MeV. That allowed us to compile the full census of
the  very bright flares in the local Universe. We have made use of the second
catalog of flaring gamma-ray sources (2FAV) based on Fermi All-Sky Variability
Analysis (FAVA) \cite{FAVA}. This catalog is a collection of gamma-ray sources
which show significantly higher (or lower) photon flux in a given time window as
compared to what could be expected from the average flux from the source
direction. The catalog is based on the Fermi observations during the first 7.4
years of the mission. 

Among the AGNs in 2FAV we selected those within the test volume $V_0$.  There
are controversial determinations of $z$ for the source \mbox{PMN~J1802-3940}:
there are a number of works with lower estimation, $z=0.296$ while in several
others the source is placed at much higher redshift,  $z=1.32$. For the sake of
robustness, we did not include the source in our analysis.
Moreover, it is possible that some AGNs could be in high state for a time span
longer than 7.4 years and therefore enter the catalog as steady sources. In
order not to miss such objects, we checked 3FGL catalog for 'superluminous'
sources within $z=0.3$. For the sources with known redshift we approximated the
isotropic equivalent radiative luminosity with
$L_\gamma \sim 4 \pi d^2 F_{100}$, $d$ is the luminosity distance to the source
assuming the standard cosmology
($H_0 = 69.6\text{km s}^{-1}\text{Mpc}^{-1}, \Omega_\mr{M} = 0.286$) and
$F_{100}$ is the energy flux above 100 MeV taken from 3FGL. Only one source with
the known $z \le 0.3$, S5~0716+71,  exceeded the threshold
(\ref{eq:maglum_thr}). This source is  also in the 2FAV catalog and we included
it in our analysis. The final list of selected AGNs consists of thirty eight
sources and is shown in table~\ref{tab:longlist}.

%
\begin{table}[tbp]
\centering
\begin{tabular}{|l|cccccc|}
\hline
Name & $z$ & $l$ & $b$  & $E_\mr{kin}$ & \% &
$L_\mr{max}$ \\
\hline
PKS 0056-572	& 0.02	& $300.9$	& $-60.1$		& 0 & 0 & 0.0021 \\
PKS 0131-522	& 0.02	& $288.3$	& $-63.9$		& 0 & 0 & 0.0084 \\
Mkn 421	& 0.03	& $179.8$	& $65.0$		& 0 & 0 & 0.11 \\
3C 120	& 0.03	& $190.4$	& $-27.4$		& 0 & 0 & 0.027 \\
Mkn 501	& 0.03	& $63.6$	& $38.9$		& 0 & 0 & 0.037 \\
1ES 1959+650	& 0.05	& $98.0$	& $17.7$		& 0 & 0 & 0.090 \\
SBS 1646+499	& 0.05	& $76.6$	& $40.1$		& 0 & 0 & 0.035 \\
3C 111	& 0.05	& $161.7$	& $-8.8$	& 0	 & 0 & 0.052 \\
AP Librae	& 0.05	& $340.7$	& $27.6$		& 0 & 0 & 0.059 \\
5BZBJ1728+5013	& 0.06	& $77.1$	& $33.5$		& 0 & 0 & 0.058 \\
PKS 0521-36	& 0.06	& $240.6$	& $-32.7$	& 0 & 0 & 0.16 \\
1H 0323+342	& 0.06	& $155.7$	& $-18.8$		& 0 & 0 & 0.11 \\
PKS 1441+25	& 0.06	& $34.6$	& $64.7$		& 0 & 0 & 0.36 \\
BL Lacertae	& 0.07	& $92.6$	& $-10.4$		& 0 & 0 & 0.35 \\
TXS 0518+211	& 0.11	& $183.6$	& $-8.7$		& 0 & 0 & 0.85 \\
PKS 2155-304	& 0.12	& $17.7$	& $-52.2$		& $6.1 \times 10^{50}$ &  0.1	 & 1.2 \\
GB6 J1542+6129	& 0.12	& $95.4$	& $45.4$		& 0 & 0 & 0.27 \\
1ES 1215+303	& 0.13	& $188.9$	& $82.1$		& 0 & 0 & 0.98 \\
ON 246	& 0.14	& $232.8$	& $84.9$		& $2.6 \times 10^{51}$ & 0.4	 &  1.5 \\
PKS 1717+177	& 0.14	& $39.5$	& $28.1$		& 0 & 0 & 0.73 \\
1ES 0806+524	& 0.14	& $166.2$	& $32.9$		& 0 & 0 & 0.50 \\
OQ 530	& 0.15	& $98.3$	& $58.3$	& 0	 & 0 & 0.45 \\
3C 273	& 0.16	& $290.0$	& $64.4$		& $6.8 \times 10^{51}$ & 1.0	 &  3.5 \\
PKS 0829+046	& 0.17	& $220.7$	& $24.3$		& $5.9 \times 10^{50}$ &  0.1	 & 1.1 \\
PKS 0736+01	& 0.19	& $217.0$	& $11.4$		& $9.2 \times 10^{51}$ & 1.4 &  3.4 \\
MG1 J021114+1051	& 0.20	& $152.6$	& $-47.4$  & $2.0 \times 10^{51}$ &  0.3	 & 1.8 \\
B2 2107+35A	& 0.20	& $80.3$	& $-8.4$		& 0 & 0 & 0.71 \\
OX 169	& 0.21	& $72.1$	& $-26.1$		& $3.1 \times 10^{51}$ & 0.5	 &  1.4 \\
1H 1013+498	& 0.21	& $165.5$	& $52.7$		& $5.3 \times 10^{51}$ & 0.8 &  2.5 \\
B2 2023+33	& 0.22	& $73.1$	& $-2.4$		& $2.7 \times 10^{53}$ &  41.4	 & 4.0 \\
PMN J0017-0512	& 0.23	& $101.2$	& $-66.6$		& $7.4 \times 10^{50}$ &  0.1	 & 1.7 \\
5BZQJ0422-0643	& 0.24	& $200.8$	& $-36.1$		& 0 & 0 & 0.91 \\
PMN J1903-6749	& 0.26	& $327.7$	& $-26.1$		& $2.0 \times 10^{51}$ &  0.3	 & 2.0 \\
PKS 0301-243	& 0.26	& $214.6$	& $-60.2$	 & $1.7 \times 10^{52}$ &  2.6	 & 9.3 \\
S2 0109+22	& 0.27	& $129.1$	& $-39.9$		& $5.4 \times 10^{52}$ &  8.2	 & 3.7 \\
GB6 J0937+5008	& 0.28	& $167.4$	& $46.7$		& $6.1 \times 10^{50}$ &  0.1	 & 1.2 \\
NVSS J223708-392137	& 0.30	& $0.6$	& $-59.6$		& $3.3 \times 10^{51}$ &  0.5	 & 1.9 \\
S5 0716+71	& 0.30	& $144.0$	& $28.0$		& $2.8 \times 10^{53}$ &  42.3	 & 11 \\
\hline
\end{tabular}
\caption{\label{tab:longlist}The list of the analyzed AGNs. $z$, $l$, $b$ are
the redshift and galactic longitude and latitude, $E_\mr{kin}$ is the kinetic energy produced in the flares in
the source, \% is the percentage of this energy in the total kinetic energy of
all flares, $L_\mr{max}$ is the maximum isotropic equivalent radiative
luminosity in the observation period, in units of 'critical luminosity'
$4 \times 10^{46} \text{erg s}^{-1}$, assuming $\Gamma=10$. All the estimates
were made with the aperture photometry method.}
\end{table}

There are different ways to estimate the gamma flux  with their own advantages
and drawbacks. First, one can use aperture photometry (AP) light curves which
are available, e.g., on the SSDC website\footnote{http://www.asdc.asi.it/}. This
simple approach  does not distinguish between source and background photons,
therefore the flux is inevitably overestimated, especially at low galactic
latitudes. Second, one can obtain maximum likelihood (ML) flux estimates for
each flare using a Fermi Science Tools routine \textit{gtlike}. This approach is
the most rigorous, but the total length of the time intervals to be analyzed
and their number makes such a task more challenging. Therefore, we solve the
task in two stages: first, calculate the kinetic energy output using AP data and
estimate the contribution of each source; second, recalculate this quantity
using ML, but only for the sources with large contribution.

At the first stage, we compute the flux from AP data obtained from SSDC and use
it to find strong flares exceeding our threshold and after that  estimate the
total  kinetic power in these flares. There are   strong correlations between
the flux in the high energy range,  0.1 -- 100 GeV and both the radiative
luminosity and kinetic luminosity of the jet. At the same time it is desirable
to use observations at even higher energies ($>1$~GeV) in order to maximally
avoid contamination  by the background photons because the point-spread function
of Fermi-LAT quickly deteriorates with decreasing energies. Our strategy is to
calculate the AP flux in the range 1 -- 100 GeV within 2 degrees from the source
and extrapolate it down to 0.1 GeV in the assumption of a flat spectrum.

The threshold luminosity given by Eq.~(\ref{eq:maglum_thr}) depends on the bulk
Lorentz factor of the jet. We used Lorentz factor $\Gamma = 10$ as a robust
benchmark value \cite{Ghisellini15, Savolainen10, Hervet16} and considered
acceleration to  energies higher than $10^{20}$ eV. These parameters set the
threshold (\ref{eq:maglum_thr}) at the level of $4 \times 10^{46}$ erg/s. We
have explicitly checked that the acceleration in this regime was not limited by
the synchrotron losses. Due to a relatively long duration of flares ($>10^5~$s)
constraints on the minimal size of the acceleration region are greatly relaxed,
which in turn allows to decrease the magnitude of needed magnetic field and,
therefore, importance  of synchrotron losses, see e.g. eqs. (1) and (2) in
\cite{Farrar2009}. After that we selected the time intervals where the estimated
luminosity exceeds the threshold. In other words, we chose the time intervals
when a given AGN can accelerate protons to energies higher than $10^{20}$~eV. As
a proxy of the (synchrotron) radiative luminosity we use the  energy flux in the
range 0.1 -- 100 GeV.

After that it was possible to estimate the total kinetic energy of these flares.
One of the tightest correlation between the isotropic-equivalent gamma-ray
luminosity $L_\gamma$ and the jet kinetic power $P_\mr{jet}$ was obtained in
\cite{Nemmen2012}:
\begin{equation}
\log P_\mr{jet} = A\log L_\gamma + B \label{eq:nemmen}
\end{equation}
where $A = 0.51 \pm 0.02$ and $B = 21.2 \pm 1.1$. We assume that, due to the 
beaming of the photons, we cannot observe  all the blazars, but only a fraction 
of order $f_b \equiv \Omega_b/2\pi \sim 1/2\Gamma^2$ where $\Omega_b$ is the 
solid angle of each of the two jets. Therefore in order to obtain the estimate 
of the total kinetic energy within the test volume, for each source we integrate
$P_\mr{jet}$ over the time intervals when the source is above the luminosity
threshold, sum up the energies from all the sources and multiply the sum by $2\Gamma^2$. We stress that while
$L_\gamma$ is the isotropic-equivalent luminosity, $P_\mr{jet}$ is not.
$P_\mr{jet}$ is an estimation of the actual kinetic power of the jet.
Division by $f_b$ in turn corrects the 'observed' power for the sources
with misaligned jets (which therefore are not observed as blazars). The kinetic
energy outputs obtained for each source is shown in table~\ref{tab:longlist}. Their sum, after multiplication by
$2\Gamma^2$, yields $ (1.3 \pm 0.2) \times 10^{56}$ erg. One
can see that only a handful of the sources can potentially contribute to the
acceleration of CRs of the highest energies. For these sources we constructed
more accurate ML light curves with the 1 week cadence in the energy range
0.1--100 GeV with the \textit{Fermipy} package \cite{Wood2017} and used them
instead of AP lightcurves. The refined kinetic energies are given in
table~\ref{tab:shortlist}. Note that the contribution from B2~2023+33 dropped
down severely because of its low galactic latitude.

\begin{table}[tbp]
\centering
\begin{tabular}{|l|ccc|}
\hline
Name & Type & $E_\mr{kin}^\mathrm{AP}$ & $E_\mr{kin}^\mathrm{ML}$\\
\hline
3C 273 & FSRQ & $6.8 \times 10^{51}$ & $7.7 \times 10^{51}$ \\
PKS 0736+01 & FSRQ & $9.1 \times 10^{51}$ & $4.5 \times 10^{51}$ \\
B2 2023+33 & BL Lac & $2.7 \times 10^{53}$ & $3.3 \times 10^{51}$ \\
PKS 0301-243 & BL Lac & $1.7 \times 10^{52}$ & $9.0 \times 10^{51}$ \\
S2 0109+22 & BL Lac & $5.4 \times 10^{52}$ & $1.7 \times 10^{52}$ \\
S5 0716+714 & BL Lac & $2.8 \times 10^{53}$ & $1.9 \times 10^{53}$ \\

\hline
\end{tabular}
\caption{\label{tab:shortlist} Refined kinetic energies for the sources
with the largest power contribution obtained using maximum likelihood approach.
The column with the AP-calculated energies is repeated for reference.}
\end{table}

The total kinetic energy which does not suffer from the drawbacks of AP any more
is $\mathrm{(4.7 \pm 1.0)} \times 10^{55} $ erg.

There is a caveat in using the empirical relation between $L_\gamma^\mr{iso}$
and $P_\mr{jet}$ from \cite{Nemmen2012}: they estimated the kinetic power from
the observations of the cavities around AGNs; these cavities were probably
produced $10^7-10^8$ years ago and may not represent the current AGN power.
However,  the empirical relation between the luminosity and kinetic power is
valid not only for AGNs, but also for GRBs, although in the latter case it should be used with some degree of caution, as it is still a matter of debate. Moreover, \cite{Lamb2017} traced the
same relation down to the luminosities of X-ray binaries  in the Milky Way during flares so
that the whole relation spans over $\sim 17$ orders of magnitude in luminosity.
This 'unified' relation has the form $\log P_\mr{jet} = \alpha \log L + \beta$
where $L$ is the collimation corrected luminosity $L=f_b L_\gamma^\mr{iso}$,
$\alpha=0.98\pm0.02$ and $\beta = 1.6\pm0.9$. If we use this relation with the
adopted value of $\Gamma=10$ instead of Eq.~(\ref{eq:nemmen}), the resulting
kinetic energy is $(7.9 \pm 1.6) \times 10^{55} $ erg. In the
following, we use the former value of $\mathrm{4.7} \times 10^{55}$ erg.

The kinetic energy is divided by the timespan  of observations (7.4 years) and
by the volume $V_{0}$ -- that gives the total kinetic emissivity. The value
obtained is $3.7 \times 10^{44}$ erg Mpc$^{-3}$ yr$^{-1}$. This should be
compared with the emissivity in UHECRs. It was shown that  the emissivity
required to reproduce the UHECR data above $10^{18}$ eV should be of the order
$10^{45} - 10^{46}$ erg Mpc$^{-3}$ yr$^{-1}$ \cite{Aloisio2017}  which
corresponds to $\sim 10^{43} - 10^{44}$ erg Mpc$^{-3}$ yr$^{-1}$ above $10^{20}$
eV. Our own calculation which follows the approach of \cite{Stecker1968} using
the photo-pion production cross-section from \cite{PDG16}
and pair production cross sections from \cite{Blumenthal1970, Chodorowski1992} gave the values of
$8.1 \times 10^{43}$ erg Mpc$^{-3}$ yr$^{-1}$ and $1.4 \times 10^{43}$
erg Mpc$^{-3}$ yr$^{-1}$ (see Appendix~\ref{sec:C}). The first value corresponds
to the UHECR intensity as observed by Telescope Array collaboration
\cite{TA_GZK}, and the second value corresponds to the intensity reported  by
the  Pierre Auger collaboration \cite{PA_spec}. Thus, the ratio of the UHECR
emissivity to the AGNs' jet kinetic power, according to our estimations, varies
from $3.8 \times 10^{-2}$ in case of the lower UHECR flux of Pierre Auger to
$2.2 \times 10^{-1}$ in case of the higher UHECR flux of Telescope Array. In
Appendix \ref{app:thr} we show how this estimation is affected by the
uncertainty of the threshold (\ref{eq:maglum_thr}).

Let us outline once again the procedure we used to obtain the kinetic emissivity of blazars.
\begin{enumerate}
    \item For each source, find the time intervals when the luminosity exceeds the threshold. We call such intervals 'flares'.
    \item Within these intervals, convert the radiative luminosity to the kinetic power via Eq.~(\ref{eq:nemmen}) and integrate it over time to obtain the kinetic energy output from the source in flaring states over the whole timespan.
    \item Sum up the energy outputs from all the sources and divide the sum by $f_b$, the beaming factor.
    \item Divide by the volume and the timespan of the observations to obtain the average kinetic emissivity of flaring AGNs.
\end{enumerate}
\section{Discussion}\label{sec:dis}
\subsection{UHECR isotropization and acceleration spectrum}\label{sec:dis:iso}
In the previous section we have calculated emissivity in UHECRs
$\mathcal{L}_{\mathrm{UHECR}}=8.1 \times 10^{43}~\mathrm{erg~Mpc^{-3}~yr^{-1}}$
or $1.4 \times 10^{43}$ erg Mpc$^{-3}$ yr$^{-1}$ from the observed flux
(reported by Telescope Array or Pierre Auger) of these particles and compared it
with the kinetic emissivity in the strong AGN flares which satisfy
Eq. (\ref{eq:maglum_thr}):
$\mathcal{L}_{\mathrm{kin}}=3.7 \times 10^{44}~\mathrm{erg~Mpc^{-3}~yr^{-1}}$.
The latter one was calculated taking into account the fact  that we can observe
only a small fraction of all flaring sources and total observed kinetic
emissivity should be multiplied by the number of unseen sources,
$n=2\pi/\Omega_b \sim 2\Gamma^2$.
The viability of the scenario crucially depends on the degree of UHECR beaming,
$\Omega_\mathrm{UHECR}$. If the accelerated UHECRs remain strongly beamed with
$\Omega_\mathrm{UHECR}\sim \Omega_b$, the UHECR emissivity
$\mathcal{L}_{\mathrm{UHECR}}$ will  be corrected for the beaming factor as
well. As after this correction $\mathcal{L}_{\mathrm{UHECR}}$ will be much
larger than  $\mathcal{L}_{\mathrm{kin}}$ it would clearly make  the model
unfeasible.

The UHECRs can be effectively isotropised during their propagation towards the
observer, increasing $\Omega_{\mathrm{UHECR}}$ up to value of order unity. The
isotropization can take place either in the  immediate vicinity of the flaring
region, which  in turn shall be very close to the AGN engine, or  in the
magnetized intracluster medium. Large values of $\Omega_{\mathrm{UHECR}}$ mean
that the observed flux of cosmic rays is generated by large number of flares
with only small fraction of them pointing at us. In \cite{Murase2012,Takami2016} it was shown that the average  degree of isotropization should be high, otherwise we should have observed much stronger degree of anisotropy;  only relative minority of sources could reside in void-like regions where somewhat  lower degree of isotropization is expected.
 
\begin{figure}[tbp]
\centering
	\includegraphics{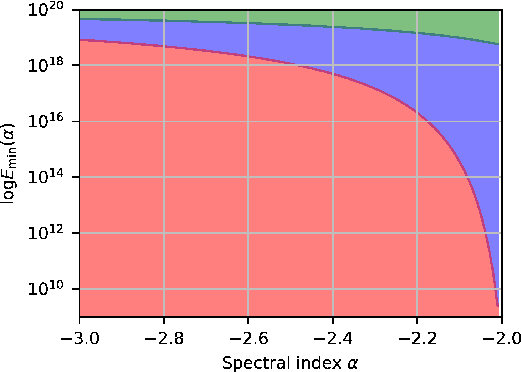}
	\caption{Constraints on the spectrum of the UHECRs  produced in  extreme AGN
	flares. The spectrum is described by a power law:
	$dN/dE\sim E^{\alpha},~E_{\mathrm{min}}(\alpha)<E<E_{\mathrm{max}}=10^{21}~$~eV.
	The green region allows to accommodate both TA and PAO data, the blue
	region is compatible to PAO data only and the red region is excluded.}
	\label{fig:spindex}
\end{figure}

Even then, the luminosity in the UHECRs with $E>10^{20}$~eV amounts to a sizable
fraction of the full kinetic  luminosity of suprathreshold AGN flares in the
local Universe, $3.8 \times 10^{-2}$ or $2.2 \times 10^{-1}$  depending on the
UHECR spectrum selected, either  PA or TA.  Nevertheless, from the point of view
of pure energetics it is not impossible that these flares can be the primary
sources of the CRs of the highest energies.  More than that, relatively large
value of the ratio of UHECR to the full kinetic luminosity allows to put
stringent constraints on the properties of the spectrum of CRs produced in
the flares. We demanded that the total amount of energy in these CRs could not
exceed one half  of the kinetic energy \cite{Berezhko1999}. The spectrum of the
UHECRs was described as a simple power-law,
$\propto E^{-\alpha},~E_\mr{min}<E<E_\mr{max}=10^{21}~$eV bounded from above and
below. The results in form of $E_\mr{min}(\alpha)$ curves are presented in the
figure~\ref{fig:spindex}. We have also checked that the results are not
considerably changed with  an increased value of $E_\mr{max}=10^{22}$~eV. It can
be seen that soft extended spectra are   excluded with a high degree of
confidence, and the spectrum of produced UHECR must be sufficiently hard and/or
narrow. A lot of models predict the very same shape for the spectrum of
particles escaping from relativistic collisionless shocks (e.g.,
\cite{Meli2013,Globus2015,Bykov2017}) or accelerated immediately in BH
magnetospheres \cite{Ptitsyna2016}.
 
Though we have used  large test volume, very strong blazar flares necessary for acceleration of UHE protons  are so rare, that  still there is rather high level of statistical fluctuations in our estimations. If we took 
S5~0716+71 for an upward fluctuation and removed its contribution as an  outlier then our estimations would drop by factor 6 that could pose severe problems for scenario in case of the TA spectrum. Low number of flare incidents in $V_0$ does not allow to evaluate  distribution of flare properties.

Naturally, a large portion of lower energies UHECRs can also be accelerated in the AGNs. There are two factors that makes it more feasible: first,  threshold luminosity becomes progressively lower  ($L_{thr}\propto E^{-2}$), which in turn makes acceleration in less extreme flares or even steady sources possible. Second, at the same time, propagation distance quickly increases which enlarge the 'active volume' and also enhance the observed flux. More detailed study of this scenario and its comparison with observed spectrum and, what is very important, mass composition at these energies will be done elsewhere.

\subsection{Multimessenger considerations}
Models with pure protonic composition can be constrained using secondary
particles -- cosmogenic neutrinos and gamma-rays which are generated in
interactions of UHECRs with background (e.g. \cite{Rachen1993}). Cosmogenic photons via multiple
$\gamma$-$e^+e^-$ conversions eventually cascade down to GeV-TeV energy range,
and can contribute considerably to isotropic diffuse gamma-ray background at
these energies.  The latest results of the Fermi-LAT \cite{Fermi_IDGRB} have
already allowed to strongly constrain  protonic models at UHECR energies around
$10^{18.2}$~eV \cite{Liu2016}. On the other hand  only models with a very strong
redshift evolution are excluded at the  highest energies \cite{Kalashev2016}.
The constraints from the cosmogenic neutrino are somewhat weaker but less
model-dependent, they are  still  compatible with protonic models, again in the
absence of strong evolution \cite{Heinze2016,Kalashev2016}. It is worth noting that if UHECRs were accelerated in immediate vicinity of AGN engine, e.g. in inner jets, some fraction of  high-energy neutrino could  have been produced by UHECRs in photo-meson interactions, which in turn makes constraints coming from cosmogenic neutrinos more stringent \cite{Murase2014} 

\subsection{UHECR source anisotropy}
Small number of observed events with $E>10^{20}$~eV  does not allow to perform
meaningful large-scale anisotropy studies, it is still impossible to reliably
distinguish between  isotropic or some anisotropic distributions. Nevertheless,
it is safe to claim that  at any given moment the number of active sources in
the $V_\mr{GZK}$ must be  larger or equal than one \cite{Troitsky2012}. We have
observed $N=6$ potential sources in $t=7.4$~years inside
$V_{0}= 10^3~V_\mr{GZK}$ volume  which gives the following estimate for the rate
of extreme UHECR producing flares in the GZK volume:
\begin{equation}
\mathcal{R}\sim \frac{N}{t}\times\frac{V_\mr{GZK}}{V_0}\Gamma^2\sim 0.1~\mathrm
{yr}^{-1}
\label{eq:rate_gzk}
\end{equation}

Number of active UHECR sources can be readily estimated as 

\begin{equation}
\mathcal{N}=\mathcal{R}t_{\mathrm{UHECR}},
\label{eq:number_gzk}
\end{equation}
where $t_{\mathrm{UHECR}}$ is the duration of observed UHECR signal. 
At first glance, given that the typical duration of gamma flares $t_{\gamma}$
are of order of  weeks\footnote{Also one cannot exclude the possibility that
individual flares can emerge during much longer interval when the source is in a
state of increased activity.}, this low  rate seems to be  at variance with the
observations. However, being charged particles, UHECRs are subject to
considerable deflection in the magnetic fields; when propagating  short pulses
are broadened by  scattering:

\begin{equation}
\tau\sim d\theta_s^2/2c,
\label{eq:scatter_time}
\end{equation}
where $d$ is the propagation path length, $\theta_s$ is the characteristic angle
of scattering. The observed duration $t_{\mathrm{UHECR}}=\tau$ can be much
longer than $t_{\gamma}$. There are two closely related causes of scattering:
first, there is a part due to the deflection in the random magnetic fields;
second, there is considerable scattering due to a finite energy width of the
pulse -- cosmic rays at different energies are deflected by different angles. 

There are several regions where scattering can possibly take place. At the very
least, a pulse of UHECRs will be deflected in the Galactic magnetic fields. At
these energies, the magnitude of scattering due to the random parts of the
galactic magnetic field  will be several tenth of degree for out of the Galactic
plane directions \cite{Pshirkov2013}, corresponding length is of order of
kpc. Also the signal will be spread by several degrees in the regular galactic
magnetic fields \cite{Pshirkov2011}. That translates to characteristic time
scales $\tau$ around several decades. It could reach more than $10^3$~yr if a
dipole field is present \cite{Murase2009}. Also UHECRs will be scattered
during inevitable isotropization in the vicinity of the source
(see Section~\ref{sec:dis:iso} above and \cite{Kalli2011,Takami2012}) -- the corresponding timescales are highly
uncertain, they could be very large if the UHECRs were scattered in the magnetic
fields of clusters, or, on the other hand, they could be insignificantly small
if the primary site of the isotropization was very close to the source.
Another sites of possible scattering include  the local filament of the Large
scale structure \cite{Yuksel2012} or the extragalactic voids \cite{Pshirkov2016,
Murase2009}. In both cases deflections are expected to be lower than 10 degrees even in the most extreme scenarios.  In the former case the  characteristic length scale is $\mathcal{O}(Mpc)$ and the  possible  time delays could be of  order
$10^{5}$~years, when in the latter case they could exceed $10^{7}$~years. That
means that at the highest energies  we will simultaneously observe flares from
at least $\mathcal{O}(10)$ sources and probably many more.
It should be stressed that we do not demand isotropization of arrival directions of CRs coming from a single source because in case of light composition at these energies   it would  have required unreasonably high Galactic magnetic fields, like $\mathcal{O}(\mu G)$   strong magnetic field with a characteristic scale exceeding hundreds of kpc. Still, much smaller expected amplitudes of deflections and corresponding delays are sufficient to  drastically increase number of simultaneously observed UHECR sources. Also a residual degree of anisotropy, like correlation with the local large scale structure is expected due to non-homogeinity of the matter distribution in the $V_\mr{GZK}$.

\subsection{Blazar models}\label{sec:dis:models}
We should emphasize some underlying assumptions which are important to our
approach, in particular to our using of Eq.~(\ref{eq:maglum_thr}). Namely, we
assume that the synchrotron and magnetic field energy densities are
approximately in equipartition. In the framework of SSC models, the ratio of
these energy densities is equal to the ratio of the inverse Compton to
synchrotron luminosity also known as Compton dominance. Although this parameter
can vary in quite a wide range from one source to another, it is shown to
increase systematically from BL Lacs to \mbox{FSRQs}, being around unity or
even smaller for the former \cite{Fossati1998,Finke2013}. In our
Table~\ref{tab:shortlist}, all the sources except 3C~273 and PKS~0736+01 are
BL~Lac objects. Moreover, the two most energetic sources in the list, S5~0716+71
and S2~0109+22 have been observed in a wide spectral range, in particular during
a high state. Multiwavelength observations  provide spectra of S5~0716+71 in the
flaring state from which one can see that the inverse Compton peak is less
significant than the synchrotron one \cite{Ahnen2018}. The flaring spectrum of
S2~0109+22 \cite{Ansoldi2018} shows two peaks of comparable magnitude. Similar
picture is seen in other BL Lac blazars, e.g. Mkn~421, Mkn~501, 1ES~1959+650
\cite{Aleksic2015, Ahnen2017, Kaur2017}. If the high-energy peak of SED is lower
than  the synchrotron one, which is quite possible taking into account the
observations cited, then using high-energy flux as a proxy, we underestimate the magnetic energy density and overestimate the threshold luminosity which makes the constraints on the UHECR
acceleration spectrum more stringent. In this sense, our assumption of energy
equipartition and the equality of the SED peaks is on the conservative side.
Note that equipartition between electron and magnetic field energy is also considered to be likely the case. One argument is that the equipartition between electrons and magnetic field provides the minimal jet power compatible with the observed emission (see \cite{Condon2016}, chapter 5 for derivation and \cite{Dermer2014}). Equipartition is obtained as a result of numerical simulations of blazar jets. For example, \cite{Sironi2015} and \cite{Petropoulou2019} found rough equipartition between magnetic field and radiating particles in the reconnection model of blazar jets.

Our considerations refer to the one zone SSC model. In regard to BL Lacs, this
is justified by observations. For example, \cite{Rani2013} report on the
significant correlation between optical and gamma variability of S5~0716+71 and
\cite{Hovatta2014} using a large sample of blazars found that optical and
gamma-ray fluxes are correlated in BL Lacs (and in FSRQs, but to a lesser extent)
which can be naturally explained in the SSC framework. On the other hand, there
are plenty of sources where one zone SSC fails to fit all the data, e.g. for
S5~0716+71. In this case more elaborated model is needed and, e.g., two zone SSC
should be invoked instead. For our purposes  we only need that the ratio of
heights of the two spectral peaks still traces the ratio of magnetic and
radiative energies. This is the case in the spine-layer model
\cite{Ghisellini2005}.

Apart from leptonic models, there is a whole class of hadronic and
lepto-hadronic models \cite{Boettcher2013} which can also be used to fit blazar
SEDs, but result in different parameter values. However, within those models, it
is difficult to relate the Poynting luminosity to some observed luminosity and
the whole analysis deserves a separate investigation.

Some underestimation of real $\mathcal{L}_\mr{kin}$ can be due to a number of
blazars without assigned redshift. There are 65 associated sources without
assigned redshift in the 2FAV catalog, 57 of them would satisfy
Eq.~\ref{eq:maglum_thr} if they were situated at $z=0.3$. In the highly unlikely
scenario  that all sources without estimated redshifts reside inside $V_0$
volume it will shift upwards our estimate of  $\mathcal{L}_\mr{kin}$ by more
than an order of magnitude and, accordingly, the estimate of energy available
for UHECR acceleration, see Appendix~\ref{app:thr}.

Finally, we have tried to evaluate how  our results depended on the exact value
of threshold (Eq. (\ref{eq:maglum_thr})) and repeated our calculations for two
bracketing cases, i.e. $L_\mr{r}\geq 10^{46}~\mathrm{erg~s^{-1}}$ and
$L_\mr{r}\geq 10^{47}~\mathrm{erg~s^{-1}}$, see Appendix \ref{app:thr} for
details.

\subsection{UHECR composition}
We have chosen the lightest, protonic, composition for the UHECRs as a limiting
scenario. From the theoretical point of view, acceleration of protons to the
highest energies is very difficult, i.e. if it is possible for some class of
sources to accelerate them, then, \textit{a fortiori}, these sources can
potentially produce UHECR nuclei with larger atomic number $A$. From the
observational point of view, the composition of the UHECRs at the highest
energies is still far from being certain: while the results of the  Telescope
Array experiment favor lighter one \cite{TA}, the ones of Pierre Auger
Observatory indicate that there is progressive increase in atomic number $A$ at
higher energies \cite{PAO}. However, due to a very low
statistics\footnote{Present method of composition studies uses data obtained by
detectors of fluorescent light from extended showers \cite{TA_FD} and these
detectors have livetime of only $10\%$} there is no information about UHECR
composition at the energies $E>10^{20}$~eV, so it is not inconceivable that we
will eventually observe decrease in $A$. This issue will be hopefully resolved
in the near future with analysis of the surface detectors data with their much
larger duty cycle. First results of the TA collaboration analysis indicate
presence of  light UHECRs in the highest accessible energy bin
($\log E=19.6-20.0$) \cite{TA_SD2018}. Also the latest results of the PAO SD
show that the gradual increase in $A$ is apparently arrested at these
energies \cite{PAO_ICRC2017}.

On the other hand, the threshold luminosity given by the equation
(\ref{eq:maglum_thr}) is not known to the great accuracy, it can be several times
higher or lower than the value we used (see Appendix~\ref{app:thr}).
This can strongly affect the conclusions
we draw from our analysis in the assumption of protonic UHECR composition.
Namely, if the threshold is significantly higher than that given by
Eq.~(\ref{eq:maglum_thr}), then the kinetic emissivity of AGNs is insufficient
to account for the observed UHECR intensity. Therefore it is reasonable to
perform our analysis in the assumption of heavier composition. One can
show that, in such case, the threshold luminosity used in our analysis lowers by
$Z_\mr{obs}^2$, the charge of the observed nuclei
squared\footnote{Eq.~(\ref{eq:maglum_thr}) contains the scaling $\sim 1/Z^2$,
i.e. the square of the primary nuclei charge. Note however that in order
for an observed particle of the mass $A_\mr{obs}$ to have the energy $E$, the primary
particle of the mass $A$ should have been accelerated to the energy
$EA/A_\mr{obs} \approx EZ/Z_\mr{obs}$. Substituting this energy
to Eq.~(\ref{eq:maglum_thr})
we obtain the scaling $\sim 1/Z_\mr{obs}^2$}. E.g. in the case of
carbon nuclei, the threshold lowers by a factor of 36 to $10^{45}$ erg/s.
In fact, at this level of gamma luminosity, we no longer need to seek for
strong flares because there is quite a number of steady sources fulfilling
the 'loosened' criterion. Namely, we found 76 blazars in 3FGL with $z \leq 0.3$ and $L \geq 10^{45}$~erg/s. Their total kinetic emissivity appears to be
$8 \times 10^{45}~\mr{erg~Mpc^{-3}~yr^{-1}}$. The estimation of the
theoretical UHECR emissivity depends on what nuclei are assumed to be
accelerated at sources: the heavier the primaries, the lower emissivity
is required to supply the observed intensity. The calculation is described
in Appendix~\ref{app:nucl}. For the PAO data we obtained
the emissivity of $3 \times 10^{44}~\mr{erg~Mpc^{-3}~yr^{-1}}$ for carbon primaries and
$5 \times 10^{43}~\mr{erg~Mpc^{-3}~yr^{-1}}$ for silicon primaries. Our
estimates show that, even for carbon,
the ratio of UHECR emissivity to the total kinetic emissivity is less than
4\% and is even lower for heavier primaries/observables. This makes the whole
scenario more feasible. Also, due to stronger deflection of UHE
nuclei in the magnetic fields and higher  number of potential sources, the
expected degree of anisotropy in this scenario is naturally lower.
\section{Conclusions} \label{sec:conclusion}
The full census of strong local gamma-ray flares at redshifts $z<0.3$ from the
Fermi-LAT data  was used in order to find ones that can possibly accelerate
protons to the highest energies $E>10^{20}$~eV and  estimate  maximal
disposable amount  of kinetic energy that can be potentially  used for UHECRs
acceleration. The estimated  kinetic emissivity is approximately one order of
magnitude higher than the UHECR  emissivity obtained  from the observations,
that makes the scenario feasible from the point of view of total energetics, if
the escape spectrum of cosmic rays is not too soft. Also,  the number of
potential sources in the Greisen-Zatsepin-Kuzmin volume is  high enough, so no
markedly anisotropic distribution of UHECRs is expected which is perfectly in
line with the current observations. Thus, the giant flares of the AGNs similar
to ones  observed  with the Fermi-LAT  can be primary sources of the
UHECRs with energies $E>10^{20}$~eV.
\appendix
\section{UHECR emissivity: protons} \label{sec:C}
Here we present the calculation of the UHECR emissivity in the local volume
needed to provide the observed UHECR flux above $10^{20}$ eV. Let $N$ be
the particle distribution function in space and energy and $E$ is the particle
energy. In general, $N$ depends on the three space coordinates so that
$\int N(\mathbf{r}, E)dVdE$ is the number of the particles inside the volume
$V$. Consider the propagation of the particles along  coordinate $s$. The
kinetic (transport) equation in one dimension is essentially the continuity
equation in the space $(s, E)$ and can be written in the form
\be
	\frac{\partial N}{\partial t} = -\frac{\partial}{\partial s}(Nv)
	- \frac{\partial}{\partial E}\left(N\frac{dE}{dt}\right) + S
\ee
where $t$ is time, $v$ is the particle velocity, $dE/dt = vdE/ds$ is the energy
loss rate, or the velocity in the energy space, and $S$ represents the source
term. For our simple estimation we will assume that the medium is uniform and
isotropic on the scale of 100 Mpc which justifies the use of one-dimensional
approach  and implies $\partial/\partial s \equiv 0$. We also assume 
stationarity on the scales of $\sim$300 Myr which leads to the relation
$\partial/\partial t \equiv 0$. Moreover, at such relatively small scale we
neglect the adiabatic losses.
These assumptions leave us with
\be
	S = \frac{\partial}{\partial E}\left(Nv\frac{dE}{ds}\right).
\ee
The energy losses are due to photo-pion production and pair production, therefore
\be
	dE/ds = -E/\lambda_\mr{attn} \label{eq:losses}
\ee
where $\lambda_\mr{attn}$ is the mean attenuation free path of a CR due to pion and pair production:
\be
    \frac{1}{\lambda_\mathrm{attn}} = \frac{1}{\lambda_\mathrm{pion}} + \frac{1}{\lambda_\mathrm{pair}}.
\ee
$\lambda_\mathrm{pion}$ can be
obtained from eq.~10 of \cite{Stecker1968} via multiplication by the CR velocity
$c$ and changing the order of integration:
\begin{equation}
	\lambda_\mr{pion} = \frac{2 \gamma^2 \hbar^3 \pi^2 c^3}{kT} 
	\times \left\lbrace\int_{\epsilon_\mr{th}'}^\infty d\epsilon' \epsilon'
	\sigma(\epsilon') K_p(\epsilon')[-\log(1 - e^{-\epsilon' / 2 \gamma k T})]
	\right\rbrace^{-1}.
\end{equation}
Here $T$ is the temperature of CMB, $\gamma$ is the Lorentz-factor of the UHECR,
$\epsilon'$ is the CMB photon energy in the UHECR rest frame,
$\epsilon_\mr{th}'$ is the pion production threshold energy, $\sigma(\epsilon')$
is the photo-pion production cross-section and $K_p(\epsilon')$ is the
interaction inelasticity, or the mean fraction of the energy lost by an UHECR in
a photo-pion production event.
$\lambda_\mathrm{pair}$ can be found in \cite{Blumenthal1970}:
\be
    \lambda_\mathrm{pair} = \frac{\pi^2 \hbar^3 c^3}{\alpha r_0^2 Z^2 (mc^2 kT)^2} \frac{E}{f(\nu)}
\ee
with $r_0$ the classical electron radius, $m$ the electron mass, $Z$ the UHECR charge, $\nu = mc^2/2\gamma kT$ and
\be
    f(\nu) = \nu^2 \int_2^\infty d\xi \phi(\xi)(e^{\nu \xi} -1)^{-1}.
\ee
The approximation for the function $\phi(\xi)$ was taken from \cite{Chodorowski1992}.
Equation (\ref{eq:losses}) is equivalent to eq.~1
of \cite{Berezinsky06}.

For ultrarelativistic particles the density and intensity are related via
\be
	I = \frac{c}{4\pi}N,
\ee
hence, we obtain
\be
	S = -4\pi\frac{\partial}{\partial E}\frac{IE}{\lambda_\mr{attn}}
\ee
which in principle expresses the spectrum of the UHECR sources given the
observed spectrum and the photo-pion production cross-section. Now the total
energetic emissivity in UHECRs can be obtained via integration of the last
equation over energies with the weight $E$:
\begin{equation}
	\int_{E_\mr{min}}^{E_\mr{max}} ESdE = -4\pi\int_{E_\mr{min}}^{E_\mr{max}} E
	\frac{\partial}{\partial E}\frac{IE}{\lambda_\mr{attn}} dE
	=\left.\frac{4\pi E^2 I }{\lambda_\mr{attn}}
	\right|^{E_\mr{min}}_{E_\mr{max}} + 4\pi \int_{E_\mr{min}}^{E_\mr{max}}
	\frac{I E}{\lambda_\mr{attn}} dE \label{eq:correct}
\end{equation}
where $E_\mr{min}=10^{20}$ eV and $E_\mr{max}=10^{21}$ eV is the assumed upper
energy limit of UHECR acceleration.
The spectrum of cosmic rays above $10^{20}$ eV as observed by the Telescope
Array collaboration can be represented in the form
\be
	I(E) = A\left(\frac{E\text{, eV}}{10^{20}}\right)^{-\gamma}
\ee
where $A = 6.25 \times 10^{-29}$ erg$^{-1}$cm$^{-2}$s$^{-1}$sr$^{-1}$ and
$\gamma = 4.6$ \cite{TA_GZK}. The spectrum in this energy range provided by the
Pierre Auger collaboration reads
\be
	I(E) = I_0 \left(\frac{E}{E_a}\right)^{-\gamma_2} \left[1
	+\left(\frac{E_a}{E_s}\right)^{\Delta\gamma}\right]
	\left[1+\left(\frac{E}{E_s}\right)^{\Delta\gamma}\right]^{-1}
\ee
where $I_0 = 3.30 \times 10^{-19}$ eV$^{-1}$km$^{-2}$sr$^{-1}$yr$^{-1}$,
$E_\mr{a} = 4.8$ EeV, $E_\mr{s} = 42$ EeV, $\gamma_2 = 2.60$ and
$\Delta \gamma = 3.14$ \cite{PA_spec}. Substituting these expressions to
(\ref{eq:correct}) we obtain for TA and PA data $8.1 \times 10^{43}$ and
$1.4 \times 10^{43}$ erg Mpc$^{-3}$yr$^{-1}$ respectively. 

\section{UHECR emissivity: heavy nuclei} \label{app:nucl}
The calculation for heavy nuclei is different from that for protons due to the fact that they lose energy by photodisintegration which means they conserve the initial Lorentz factor while decreasing the energy. One can see from figure~2 of \cite{Aloisio2013a} that, in the energy range of interest ($>10^{20}$ eV) losses due to pair production are less than the photodisintegration losses. Only at about $A=40$ pair production starts to dominate at $\sim 10^{20}$ eV. We therefore neglected losses due to pair production. Then the transport equation for the nuclei of the sort $i$ is simply
\be
\frac{\partial N_i(\Gamma)}{\partial t} = -\frac{N_i(\Gamma)}{\tau_i} + S_i(\Gamma) = 0
\ee
where the term $N_i/\tau_i$ represents particle losses due to photodisintegration, $\tau_i$ being the mean time between disintegration events, and the term $S_i$ represents the sources. For the primary nuclei, the source term is a free parameter. For the rest of the nuclei the source term is equal to the loss term of the previous nucleus, i.e. $S_i = N_{i+1}/\tau_{i+1}$ so that the density of the nuclei of the mass $A$ can be obtained from the corresponding equation using the solution of the equation for the nuclei of the mass $A+1$. Under our assumptions, there is no loss term for protons. Moreover, for primary nuclei heavier than carbon accelerated to $10^{21}$~eV at maximum, protons can only have energies less than $10^{20}$~eV which is out of the range of interest of the present study. The primary source term assumed to obey the power law is adjusted so that the densities sum up to the observed quantity:
\be
\sum_i N_i(E) = \frac{4\pi}{c}I(E)
\ee
We assumed acceleration in the energy range $10^{20} - 10^{21}$ eV. Even for carbon, secondary protons appear to be outside this energy range and we did not take them into account. We used the data from PAO and obtained, for primary carbon, silicon and iron nuclei, the emissivities of $3 \times 10^{44}$, $5 \times 10^{43}$ and $4 \times 10^{42}~\mr{erg~Mpc^{-3}~yr^{-1}}$ respectively, although the value for iron is underestimated due to the neglect of pair production.

\section{Uncertainty of threshold luminosity and unknown redshifts}
\label{app:thr}
The threshold (\ref{eq:maglum_thr}) is not a well-defined quantity and its real
value can deviate to a certain extent from the adopted value. We estimated the
total kinetic luminosity as described in Section  \ref{sec:met} with the
threshold luminosities of $10^{46}$ erg/s and $10^{47}$ erg/s.
The values obtained are $1.2 \times 10^{45}$ erg Mpc$^{-3}$yr$^{-1}$ and
$1.3 \times 10^{44}$ erg Mpc$^{-3}$yr$^{-1}$. The latter value is rather close
to the inferred UHECR emissivity which can make this scenario problematic.

Next, we note that in 2FAV catalog there are 65 sources with unknown redshifts.
If we place them at the boundary of the test volume ($z = 0.3$), then 57 of them
exceed the threshold (\ref{eq:maglum_thr}) at some times. Then we can calculate
the 'hypothetical' contribution of these sources to the overall kinetic
luminosity. This contribution is equal to $7.1 \times 10^{45}$ erg
Mpc$^{-3}$yr$^{-1}$ (we excluded the sources within \ang{5} from the galactic
plane). Obviously, this value makes the scenario more feasible, however we
stress that we considered a highly unlikely case when all the sources with
unknown $z$ reside at $z=0.3$ thus making the largest possible contribution to
the total kinetic emissivity. Moreover, this calculation is made with aperture
photometry and thus is an overestimation.

Finally we note that in the 3FGL catalog there are also a number of sources with
unknown redshift. Analogously, we 'put' them at the distance of $z=0.3$ and
estimate their isotropic equivalent luminosity as $4\pi d^2 F_{100}$ (see
Section \ref{sec:met}). We find that in such a case no source with unknown $z$
exceeds the threshold (\ref{eq:maglum_thr}).
\acknowledgments
This work  was  supported by RSF research
project No. 14-12-00146. The authors want to thank Andrei Gruzinov, Oleg
Kalashev and Sergey Troitsky for fruitful discussions. The work was supported by the
foundation for the Advancement of Theoretical Physics and Mathematics ``Basis''.
The authors acknowledge the support from the Program of development of
M.V. Lomonosov Moscow State University (Leading Scientific School
'Physics of stars, relativistic objects and galaxies').
The analysis is based on data and software
provided by the Fermi Science Support Center (FSSC). This research has made use
of NASA's Astrophysics Data System.

\bibliographystyle{JHEP}
\bibliography{Nizamov_Pshirkov}
\end{document}